
  \magnification=1200
  \tolerance=10000
  \baselineskip=24truept
  \parindent=1.truecm
  \def\ref{\par\noindent\hangindent 20pt}
  
  \def\mincir{\raise -2.truept\hbox{\rlap{\hbox{$\sim$}}\raise5.truept
  \hbox{$<$}\ }}
  \def\magcir{\raise -2.truept\hbox{\rlap{\hbox{$\sim$}}\raise5.truept
  \hbox{$>$}\ }}
  \def\gr{\kern 2pt\hbox{}^\circ{\kern -2pt K}} 
  
  \def\asymp{\raise -4.3truept\hbox{$ \ \widetilde{\phantom{xy}} \ $}}
  \null
  \centerline{{\bf THE GRAVITATIONAL--WAVE CONTRIBUTION TO CMB ANISOTROPIES}}
  \medskip
  \centerline{{\bf AND THE AMPLITUDE OF MASS FLUCTUATIONS FROM COBE RESULTS}}
  \bigskip
  \noindent
  \centerline{\bf Francesco Lucchin}
  \medskip
  \centerline{Dipartimento di Astronomia, Universit\`a di Padova,}

  \centerline{vicolo dell'Osservatorio 5, I--35122 Padova, Italy}
  \bigskip
  \centerline{\bf Sabino Matarrese}
  \medskip
  \centerline{Dipartimento di Fisica {\it Galileo Galilei}, Universit\`a di
  Padova,}

  \centerline{via Marzolo 8, I--35131 Padova, Italy}
  \bigskip
  \centerline{\bf Silvia Mollerach}
  \medskip
  \centerline{NASA / Fermilab Astrophysics Center,}

  \centerline{P.O. Box 500, Batavia, Illinois 60510, USA}
  \bigskip
  \centerline{July 1992}
  \vskip1.truecm
  \baselineskip=18truept
  \noindent
  {\bf Abstract} A stochastic background of primordial gravitational waves
  may substantially contribute, via the Sachs--Wolfe effect, to the
large--scale
  Cosmic Microwave Background (CMB) anisotropies recently detected by COBE.
  This implies a {\it bias} in any resulting determination of the
  primordial amplitude of density fluctuations. We consider the constraints
  imposed on $n<1$ (``tilted") power--law fluctuation spectra,
  taking into account the contribution from both scalar and tensor waves,
  as predicted by power--law inflation. The
  gravitational--wave contribution to CMB anisotropies generally reduces the
  required {\it rms} level of mass fluctuation, thereby increasing the linear
  {\it bias parameter}, even in models where the spectral index is
  close to the Harrison--Zel'dovich value $n=1$. This ``gravitational--wave
  bias" helps to reconcile the predictions of CDM models with
  observations on pairwise galaxy velocity dispersion on small scales.
  \medskip
  \noindent {\it Subject headings:} cosmic background radiation ---
  cosmology --- early universe --- galaxies: formation
  \baselineskip=24truept
  \vfill
  \eject

  The recent detection of large angular scale CMB anisotropies by the
  COBE satellite (Smoot {\it et al.} 1992) opens a window to the understanding
  of the physics of the early universe:
  in particular, it provides strong constraints on models for the origin
  of primordial perturbations. Inflation is probably the simplest
  and most motivated of such models: perturbations are
  generated in a causal way by zero--point quantum fluctuations which are
  then magnified by the accelerated universe expansion to cosmologically
  observable scales.
  The determination of the {\it rms} fluctuation amplitude
  consistent with the COBE measurements in the frame of various galaxy
formation
  scenarios (e.g. Wright {\it et al.} 1992; Efstathiou, Bond \& White, 1992;
  Schaefer \& Shafi 1992)
  has however shown that a quite high fluctuation level is
  required, which, in the standard Cold Dark Matter (CDM) scenario
  causes excessive small--scale power. In fact, a relevant quantity for
  all galaxy formation scenarios is the linear {\it bias parameter}, defined as
  the inverse of the {\it rms} linear mass fluctuation on a sphere of
  $8~h^{-1}$ Mpc ($h$ is the Hubble constant in units of
  $100$ km sec$^{-1}$ Mpc$^{-1}$; we take $h=0.5$):
  the COBE results imply $b \approx 0.8$, for the scale--invariant ($n=1$) CDM
  case. Low bias (i.e. more evolved) CDM models lead to better agreement with
  observations on large--scale flows (e.g. Bertschinger {\it et al.} 1990),
  but imply an excess of velocity dispersion on small scales when compared to
  observations on pairwise galaxy velocity dispersion in the CfA redshift
survey
  (Davis \& Peebles 1983).
  Also, the slope of the galaxy two--point function, determined in numerical
  simulations, becomes too steep. Both of these drawbacks can be
  alleviated by resorting to a velocity bias (e.g. Couchman \& Carlberg 1992).

  The above determination of $b$ is however only valid if the large
  angular scale temperature anisotropies, detected by COBE, are totally
  due to density perturbations (scalar modes), which perturb the
  last--scattering surface via the Sachs--Wolfe effect (Sachs \& Wolfe
  1967). However, a stochastic background of primordial gravitational
  waves (tensor modes) originated during inflation also contributes to
  this effect. A rough estimate of the anisotropies originated by scalar
  perturbations is $({\delta T \over T})_S \sim {1 \over 5} H \delta
  \varphi / {\dot \varphi}$, where $\varphi$ is the inflaton field,
  $\delta \varphi$ its fluctuation and $H$ the Hubble constant during
  inflation; these quantities have to be evaluated at the time when the
  scales relevant to the large--scale CMB fluctuations crossed the
  Hubble--radius, i.e. about $60$ e--foldings before the end of the
  inflationary expansion. The anisotropy originated by tensor
  perturbations is $({\delta T \over T})_T \sim {\kappa\over 2} \delta
  \varphi$, with
  $\kappa \equiv \sqrt {8 \pi G}$. Thus, we can easily obtain an approximate
  measure of their relative contribution by
  $({\delta T \over T})_S /({\delta T \over T})_T
  \sim {2 \over 5} H / \kappa {\dot \varphi} \vert_{HC}$. During inflation
  $H^2 = {\kappa^2 \over 3} (V(\varphi)+ {1\over 2} {\dot \varphi}^2)$,
  where $V(\varphi)$ is the effective inflaton potential. In
  many models, such as chaotic (Linde 1983) or new inflation
  (Linde 1982; Albrecht \& Steinhardt 1982), a slow--rollover approximation
  holds, ${\dot \varphi}^2 \ll V(\varphi) \approx 3H^2/\kappa^2$; in these
cases
  the contribution of tensor modes to $\delta T / T$ is much
  smaller than that due to scalar ones. However, in other models, such as
  power--law inflation (Abbott \& Wise 1984a; Lucchin \& Matarrese 1985), a
  slow--rollover approximation is not necessarily
  required and one can have ${\dot \varphi}^2 \sim V(\varphi)$.
  More in general, the minimal requirement on the inflaton dynamics
  is that it should lead to accelerated universe expansion,
  which implies ${\dot \varphi}^2/V(\varphi)\equiv
  \varepsilon <1$ and $({\delta T \over T})_S /({\delta T \over T})_T
  \sim {2\over 5}\sqrt{{1\over 6}+{1\over 3\epsilon}} > {\sqrt{2}\over 5}$. It
  should be pointed out, however, that the possibility to ascribe most
  of the  $\delta T/T$ signal to gravitational waves (Krauss \& White 1992)
  is restricted to models where the fluctuation spectrum is
  non--scale--invariant.

  To be more specific, let us consider the power--law inflation case, which
  has the advantage of being fully analytically tractable. In this case
  the universe expansion factor reads $a(t)=a_\star[1 +
  (H_\star/p)(t-t_\star)]^p \sim t^p$ (where $a_\star$ and $H_\star$
  refer to an arbitrary time $t_\star$ during inflation), with
  $p > 1$, and the inflaton field is assumed to have an exponential
  potential (Lucchin \& Matarrese 1985),
  $V(\varphi) \propto \exp(-\lambda \kappa \varphi)$, with
  $0<\lambda =\sqrt{2/p}<\sqrt{2}$. Such an exponential potential also
  describes the dynamics of extended inflation models (e.g. La \& Steinhardt
  1989; Kolb, Salopek \& Turner 1990). Moreover, for any inflation model
  where one scalar field rolls down a smooth potential,
  the evolution during the small range of e--foldings
  relevant for large--scale CMB anisotropies can be approximated by
  a power--law; thus, our results have a quite general validity.
  In such a case, the resulting power--spectrum of density perturbations at
  Hubble--radius crossing is proportional to $k^{2\alpha-3}$, with
  $\alpha=1/(1-p)<0$. A stochastic background of gravitational waves is also
  produced, with the same spectral behaviour at Hubble--radius crossing. In
  what follows we shall parametrize both of these spectra by the index
  $n \equiv 2 \alpha + 1=(p-3)/(p-1)<1$, which for
  scalar modes (but not for tensor ones!) gives the spectral slope on
  constant time hypersurfaces before recombination, i.e. the so--called
  primordial spectral index. Note that, the limit $p \to \infty$,
  corresponding to the de Sitter case, gives $\alpha \to 0$, or
  $n \to 1$, i.e. the Harrison--Zel'dovich fluctuation spectrum.
  For a general value of $p$ (or $n$) one can obtain the estimate
  $({\delta T \over T})_S / ({\delta T \over T})_T
  \sim {1 \over 5} \sqrt{2(3-n)\over 1-n} = {\sqrt{2 p} \over 5}$;
  thus, for values of $n$ not too far from unity, as required by the COBE
  results, we conclude that gravitational waves may make a significant
  contribution to large angular scale CMB anisotropies.

  Let us now provide a more detailed analysis of the problem. We can
  perform the usual expansion of temperature fluctuations in spherical
  harmonics ${\delta T \over T}(\theta,\phi) = \sum_{\ell,m} a_{\ell m}
  Y_{\ell m}(\theta,\phi)$, where the multipole coefficients take
  independent contributions from both scalar and tensor modes: $a_{\ell
  m} = a_{S,\ell m} + a_{T,\ell m}$. Note that, even though we wrote
  both the scalar and tensor modes as being proportional to the same
  field fluctuation $\delta \varphi$, they actually refer to independent
  quantum field fluctuations, namely the inflaton and one polarization
  state of the graviton, which simply have the same {\it rms} value.
  The squared multipole amplitudes $a_\ell^2 \equiv \sum_m \vert a_{\ell
  m} \vert^2$ have expectation values $\langle a_\ell^2 \rangle =
  \langle a_\ell^2 \rangle_S + \langle a_\ell^2 \rangle_T$, coming from
  both scalar and tensor perturbations. In the simplest case that both
  the inflaton and the graviton fluctuations have random phases, the
  multipoles $a_\ell$ are Rayleigh distributed in $2\ell + 1$
  ``dimensions" (Abbott \& Wise 1984b; Fabbri, Lucchin \& Matarrese
  1987), with {\it cosmic variance} $\sqrt{2/(2\ell + 1)} \langle
  a_\ell^2 \rangle$.
  The result for the scalar case is, in a flat universe,
  $\langle a_\ell^2 \rangle_S = {(2\ell + 1) \over 9}
  \int_0^\infty {d k \over k} \Delta_\Phi(k) j_\ell^2(2k/H_0)$
  (e.g. Bond \& Efstathiou 1987),
  where $\Delta_\Phi(k) \equiv {k^3 \over 2 \pi^2}
  {\cal P}_\Phi(k)$ is the power per
  logarithmic wavenumber of the peculiar gravitational potential $\Phi$ and
  ${\cal P}_\Phi(k)$ its power--spectrum.
  In the power--law inflation case this relation can be
  analytically integrated (e.g. Fabbri, Lucchin \& Matarrese 1987; Lyth \&
  Stewart 1992) to give
  $$
  \langle a_\ell^2 \rangle_S = {(2 \pi)^4 \over 25} {3-n \over 1-n} G
  (2 \ell +1) \biggl({H_0 \over 2}\biggr)^{n-1}
  {C(n) \Gamma(3-n) \Gamma (\ell + (n-1)/2) \over 2^{2-n} \Gamma^2(2-n/2)
  \Gamma(\ell -(n-5)/2)},
  \eqno(1)
  $$
  where the factor $C(n) \equiv {1 \over \pi^4 2^{n+1} a_\star^2}
  ({2 a_\star H_\star \over 3-n})^{3-n} \Gamma^2(2-n/2)$
  is related to the power--spectrum of the inflaton
  by ${\cal P}_\varphi(k) = 4\pi C(n) k^{n-1}$.
  The constants $a_\star$
  and $H_\star$ could be easily related to physical observables by
  matching the inflationary kinematics to the subsequent radiation and matter
  dominated eras. For the gravitational--wave contribution the result is
  (e.g. Abbott \& Wise 1984a; Fabbri, Lucchin \& Matarrese 1986)
  $$
  \langle a_\ell^2 \rangle_T = 144 ~\pi^5 G
  (2 \ell + 1){(\ell + 2)! \over (\ell -2)!} \biggl({H_0 \over 2}\biggr)^{n-1}
  C(n) \int_0^\infty dk k^{n-2} I_\ell^2(k),
  \eqno(2)
  $$
  where
  $$I_\ell(k) \equiv \int_{k\eta_E/\eta_0}^k d y {J_{\ell + 1/2}(k-y)
  \over (k- y)^{5/2}} {J_{5/2}(y) \over y^{3/2}},$$
  and $\eta_E$ and $\eta_0$ are the conformal time at the recombination
  and at the present epoch.
  In this case, the integration must be numerically performed.
  However, in the range of values of $n$ of interest for us, there is a
  nice property of tensor multipoles (e.g. Fabbri, Lucchin \& Matarrese
  1987) which makes it possible to relate them to the scalar ones in a
  simple manner. By numerically integrating
  $\langle a_\ell^2 \rangle_T$ in Eq.(2) one can show that the ratio
  $D_\ell(n)\equiv \langle a_\ell^2 \rangle_T /\langle a_\ell^2 \rangle_S$,
  for $\ell>2$, is independent of $\ell$ to a very good approximation,
  $D_\ell(n) \approx D(n)$.
  From the plots of Figure 1 this can be seen to hold in the
  spectral range $0.5 ~\mincir n < 1$, with better than $10\%$ accuracy.
  Figure 1 also shows that, for $n ~\mincir 0.8$,
  gravitational waves give the main contribution to the multipoles,
  while density perturbations dominate for larger $n$ values. The ratio
  of the tensor to the scalar contribution to the quadrupole
  is larger than that due to the higher order multipoles. In the high
  $\ell$ limit ($\ell \gg 1$) it is also possible to obtain an
  approximate asymptotic form for the tensor multipoles, along
  the lines of the computation by Starobinskii (1985) for $n=1$.
  Using Eq.(12) of Starobinskii as an approximation for $I_\ell^2(k)$ we
  obtain
  $$
  \langle a_\ell^2 \rangle_T  \approx 288 \pi^5 C(n) F(n)
  \biggl({H_0 \over 2}\biggr)^{n-1} \biggl(\ell +{1 \over 2}\biggr)^{n-2},
  \eqno(3)
  $$
  with
  $$
  F(n)\equiv \int_0^1 dx \biggl[{2\over 9\pi^2}(1-x^2)^{5-n\over 2}
  \biggl(1-{3\over2}x^2-{3\over 4}(1-x^2)x \ln\biggl({1+x\over 1-x}\biggr)
  \biggr)^2+{x^2 \over 8}(1-x^2)^{9-n\over 2}\biggr],
  \eqno(4)
  $$
  which shows the same asymptotic dependence on $\ell$ as the scalar
  multipoles. The numerical value of the ratio of the tensor to the
  scalar components is $5\%$ larger than those plotted in Figure 1
  for $\ell=30$, for values of $n > 0.5$, and $10\%$ larger, for values of
  $n$ between 0.5 and 0. We can then write any $\ell>2$ multipole in terms
  of the scalar contribution only, $\langle a_\ell^2 \rangle
  = \langle a_\ell^2 \rangle_S (1+D(n))$.
  This makes it easier to obtain the bounds imposed by the COBE determination
  of the angular correlation function $C(\theta)$ of temperature fluctuations
  to the complete {\it scalar + tensor} case.
  After dipole and quadrupole subtraction, one has $C(\theta) = (1+D(n))
  \sum_{l>2} (\Delta T_\ell^2)_S W^2(\ell) P_\ell(\cos \theta)$, where
  $(\Delta T_\ell^2)_S = (T_0^2/4\pi) \langle a_\ell^2 \rangle_S$, with
  $T_0=2.735 \pm 0.006$ K the mean temperature of the CMB radiation
  (Mather {\it et al.} 1990) and $W(\ell)=
  \exp[-(1/2)(\ell(\ell+1)/17.8^2)]$ the appropriate filter function for
  the DMR experiment (Smoot {\it et al.} 1992).
  Thus considering also the gravitational--wave contribution does not
  affect the best--fit on the primordial spectral index
  $n=1.15^{+0.45}_{-0.65}$ (Smoot {\it et al.} 1992, Wright {\it et al.}
  1992). This corresponds to a limit on the scale factor expansion power
  $p>5$ and on the scalar field potential coupling constant $\lambda <
  \sqrt{2/5}$. It affects, instead, the best--fit on the amplitude,
  since the total {\it rms}--quadrupole--normalized amplitude
  $Q_{rms-PS}= 16.3 \pm 4.6 ~\mu$K can now be written as
  $$
  Q_{rms-PS} = (Q_{rms-PS})_S \sqrt{1+D(n)}.
  \eqno(5)
  $$

  The normalization of primordial density fluctuations required to fit
  the COBE data gets modified by the same factor. In particular one gets
  $b(n)=b_0(n) \sqrt{1+ D(n)}$, where $b_0$ represents the same quantity
  calculated disregarding the gravitational--wave contribution to CMB
  fluctuations.
  Figure 2 shows the ratio $b/b_0$ as a function of $n$; notice that
  this ``gravitational--wave bias" is independent of the galaxy
  formation scenario, i.e. of the transfer function. A very good fit is
  $b/b_0 \approx \sqrt{14 - 12n \over 3 - n}$. The value of $b_0(n)$ in
  the frame of the CDM scenario, taking into account the (best--fitted)
  amplitude and the errors given by the COBE results on $C(\theta)$, can
  be easily derived by properly scaling the fits by Vittorio, Matarrese
  \& Lucchin (1988) [their Eqs.(23) and (24)]: one obtains $b_0(n)
  \approx 0.75 \times 10^{1.285(1-n)} (1 \pm 0.3)$. From the COBE
  detection of the $10^\circ$ anisotropy, $\sigma_{Sky}(10^\circ)=30 \pm
  5~\mu$K, we similarly get $b_0(n) \approx 0.82 \times 10^{1.15(1-n)}
  (1 \pm 0.2)$, in good agreement with Adams {\it et al.} (1992).
  By taking the weighted (by inverse variance) average of the two
  determinations and considering the gravitational--wave correction we
  finally get the estimate
  $$
  b(n) \approx 0.80 \sqrt{14 -12 n \over 3 - n} 10^{1.20(1-n)} (1 \pm 0.
  17).
  \eqno(6)
  $$

  As it is clear from Figure 1, gravitational waves give an even larger
  contribution to the COBE quadrupole detection, $Q_{rms}=13 \pm 6 ~\mu$K.
  In such a case we obtain $\langle a_2^2 \rangle_T / \langle a_2^2
  \rangle_S \approx {13(1-n) \over 3 - n}$ and, with the CDM
  transfer function, $b(n) \approx 0.94 \sqrt{16 -14 n \over 3 - n}
  ~10^{1.285(1-n)} (1 \pm 0.46)$, which, because of the higher cosmic
  variance, is affected by quite a large error bar.

  Let us consider some examples. From Eq.(6), by taking $n=0.8$, we get
  $b \approx 2.1$; even a value of the spectral index quite close to the
  Harrison--Zel'dovich one, such as $n=0.9$, involves a remarkable
  correction by gravitational waves ($\approx 23\%$), leading to the final
  estimate $b \approx 1.3$.

  Let us also notice that, given the inflationary model, the COBE data
  also provide a {\it determination} of the value of the Hubble constant
  at the time when the largest observable scale left the horizon, i.e.
  about $60$ e--foldings before the end of inflation. For instance, for
  a power--law inflation with $p = 11 $ ($n= 0.8$), $H_{60} \simeq 1.48
  \times 10^{-4} m_P (1 \pm 0.28)$, with $m_P$ the Planck mass.

  An important result of the present analysis is then the possibility to
  increase the estimate of the bias level: this implies less evolution
  of the considered cosmological models, thus lowering the amplitude of
  pairwise velocities on small scales. The tilted ($n<1$) CDM models
  considered here provide a natural solution to the lack of power on
  large scales and excess power on small scales of a CDM model with
  $n=1$ and $b \approx 1$. These non--standard CDM models have been
  analyzed by many authors. Vittorio, Matarrese \& Lucchin (1988) showed
  that they imply better agreement with large--scale drifts and with the
  cluster--cluster correlation function. Tormen, Lucchin \& Matarrese
  (1992; see also Tormen {\it et al.} 1992) explored in wider detail the
  advantages of these models in reproducing the large--scale peculiar
  velocity field traced by optically selected galaxy samples. A good fit
  of the angular correlation function of galaxies in the APM catalog is
  obtained by Liddle, Lyth \& Sutherland (1992), with a $n \approx 0.5$
  CDM model. More recently, Adams {\it et al.} (1992) have considered
  various cosmological constraints, while Cen {\it et al.} (1992) have
  run numerical simulations of $n \approx 0.7$ CDM models; however, they
  fix the normalization by fitting the COBE data without the
  tensor--wave contribution. This fact implies an overestimate of the
  small--scale power leading to a residual excess of pairwise velocity
  dispersion. The increase of the COBE determined biasing factor,
  resulting from our analysis, gives then an even stronger support to
  tilted CDM models.

  While completing this Letter two preprints have circulated that report
  on independent work on similar problems in the frame of various
  inflationary models (Salopek 1992; Davis {\it et al.} 1992). By a
  best--fit of the correlation function, Salopek derives the lower bound
  $p ~\magcir 11$ ($n ~\magcir 0.8$) in order to get an acceptable
  biasing factor. The Davis {\it et al.} analysis is mostly based on
  fitting the quadrupole amplitude. Our conclusions are fully consistent
  with their results.
  \bigskip
  \noindent
  {\bf Acknowledgments} We want to thank E. Roulet for his kind
  help with the numerical computations. This work was partially supported
  by Italian MURST, by Fondazione Angelo della Riccia and by DOE and NASA
  (Grant NAGW-2381) at Fermilab.

  {\bf Note added} The scalar contribution to the multipoles is here
  calculated through the leading term in the Sachs--Wolfe formula, as
  also done by Smoot {\it et al.} in fitting the COBE correlation
  function. However, the next to leading terms (e.g. Abbott \& Wise
  1984b; Fabbri {\it et al.} 1987) are no more negligible for $\ell >
  15$ and contribute to the smaller angles probed by COBE. Although this
  effect is not important for the present analysis, it should be
  properly taken into account in fitting the COBE autocorrelation
  function, especially when the experimental sensitivity will be
  increased. We acknowledge Paul Steinhardt for driving our attention
  to this point.

  \vfill\eject

  \noindent
  {\bf References}
  \medskip

  \ref {Abbott, L.F. \& Wise, M.B. 1984a, Nucl. Phys., B244, 541.}

  \ref {Abbott, L.F. \& Wise, M.B. 1984b, ApJ, 282, L47.}

  \ref {Adams, F.C., Bond, J.R., Freese, K., Frieman J.A. \& Olinto, A.V.
  1992, CITA preprint.}

  \ref {Albrecht, A. \& Steinhardt, P.J. 1982, Phys. Rev. Lett., 48, 1220.}

  \ref {Bertschinger, E., Dekel, A., Faber, S.M., Dressler, A. \& Burstein, D.
  1990, ApJ, 364, 370.}

  \ref {Bond, J.R. \& Efstathiou, G. 1987, MNRAS, 226, 655.}

  \ref {Cen, R., Gnedin, N.Y., Kofman, L.A. \& Ostriker, J.P. 1992,
  Princeton preprint.}

  \ref {Couchman, H.M.P. \& Carlberg, R.G. 1992, ApJ, 389, 453.}


  \ref {Davis, R.L., Hodges, H.M., Smoot, G.F., Steinhardt, P.J. \& Turner,
  M.S. 1992, preprint Fermilab-Pub-92/168-A.}

  \ref {Davis, M. \& Peebles, P.J.E. 1983, ApJ, 267, 465.}

  \ref {Efstathiou, G., Bond, J.R. \& White, S.D.M. 1992, preprint.}

  \ref {Fabbri, R., Lucchin, F. \& Matarrese, S. 1986, Phys. Lett., B166, 49.}

  \ref {Fabbri, R., Lucchin, F. \& Matarrese, S. 1987, ApJ, 315, 1.}


  \ref {Kolb, E.W., Salopek, D.S. \& Turner, M.S. 1990, Phys. Rev. D, 42,
3925.}

  \ref {Krauss, L. \& White, M. 1992, Yale preprint YCTP-P15-92.}

  \ref {La, D. \& Steinhardt, P.J. 1989, Phys. Rev. Lett., 62, 1316.}

  \ref {Liddle, A.R., Lyth, D.H. \& Sutherland, W. 1992, Phys. Lett., B279,
244}

  \ref {Linde, A.D. 1982, Phys. Lett., B108, 389.}

  \ref {Linde, A.D. 1983, Phys. Lett., B129, 177.}

  \ref {Lucchin, F. \& Matarrese, S. 1985, Phys. Rev. D, 32, 1316.}

  \ref {Lyth, D.H. \& Stewart, E.D. 1992, Phys. Lett., B274, 168.}

  \ref {Mather, J.C. {\it et al.} 1990, ApJ, 354, L37.}

  \ref {Sachs, R.K. \& Wolfe, A.M. 1967, ApJ, 147, 73.}

  \ref {Salopek, D.S. 1992, DAMTP preprint.}

  \ref {Schaefer, R.K. \& Shafi, Q. 1992, preprint BA-92-28.}

  \ref {Smoot, G.F. {\it et al.} 1992, ApJL, in press.}

  \ref {Starobinskii, A.A., 1985, Sov. Astr. Lett., 11, 133.}


  \ref {Tormen, G., Lucchin, F. \& Matarrese, S. 1992, ApJ, 386, 1.}

  \ref {Tormen, G., Moscardini, L., Lucchin, F. \& Matarrese, S. 1992,
  in preparation.}

  \ref {Vittorio, N., Matarrese, S. \& Lucchin, F. 1988, ApJ, 328, 69.}

  \ref {Wright, E.L. {\it et al.} 1992, ApJL, in press.}


  \baselineskip=24truept
  \centerline
  {\bf Figure captions}
  \bigskip
  \noindent
  {\bf Figure 1} The ratio of the tensor to the scalar contribution to the
  CMB multipoles, up to order $\ell=30$, as a function of the primordial
  spectral index $n$.
  \bigskip

  \noindent
  {\bf Figure 2} The gravitational--wave correction to the linear bias,
  $b(n)/b_0(n)=\sqrt{1+D(n)}$, as a function of the primordial spectral
  index $n$.

  \vfill\eject
  \bye